\documentclass[twocolumn,floats,prb,aps,unsortedaddress,superscriptaddress]{revtex4-1}
\usepackage[dvips]{graphicx}
\usepackage{amsfonts}
\usepackage{amsmath}
\usepackage{amssymb}
\usepackage{exscale}
\usepackage{eufrak}
\usepackage{afterpage}
\usepackage{upgreek}
\usepackage{color}	%% FOR COLORED NOTES IN DRAFT. COMMENT THIS LINE FOR FINAL VERSION 
\usepackage{braket}	%% For nice BRAKET notation
\usepackage{multirow}	%% For columns spanning multiple rows in Tables
\usepackage{subfigure}
\usepackage{ifpdf}
\usepackage{epstopdf}

\usepackage[normalem]{ulem}    %% To use \uline for underlining, 
\begin{document}

\title{
Quantum interference effects of out-of-plane confinement
\\
on two-dimensional electron systems in oxides
}

\author{A.~F.~Santander-Syro}
\email{andres.santander-syro@u-psud.fr}
% \affiliation{Universit\'e Paris-Saclay, CNRS, CSNSM-Centre de Sciences Nucl\'eaire 
%			et de Sciences de la Mati\`ere, 91405, Orsay, France}
\affiliation{Universit\'e Paris-Saclay, CNRS,  Institut des Sciences Mol\'eculaires d'Orsay, 
			91405, Orsay, France} 

\author{J.~Dai}
\altaffiliation[Present affiliation: ]{Institute of Physics and Lausanne Centre for Ultrafast Science (LACUS), 
			\'Ecole Polytechnique F\'ed\'erale de Lausanne, CH-1015 Lausanne, Switzerland}
\affiliation{Universit\'e Paris-Saclay, CNRS,  Institut des Sciences Mol\'eculaires d'Orsay, 
			91405, Orsay, France}
% \affiliation{Universit\'e Paris-Saclay, CNRS, CSNSM-Centre de Sciences Nucl\'eaire 
%			et de Sciences de la Mati\`ere, 91405, Orsay, France}

\author{T.~C.~R\"odel}
\affiliation{Universit\'e Paris-Saclay, CNRS,  Institut des Sciences Mol\'eculaires d'Orsay, 
			91405, Orsay, France}
% \affiliation{Universit\'e Paris-Saclay, CNRS, CSNSM-Centre de Sciences Nucl\'eaire 
% 			et de Sciences de la Mati\`ere, 91405, Orsay, France}
\affiliation{Synchrotron SOLEIL, L'Orme des Merisiers, Saint-Aubin-BP48, 91192 Gif-sur-Yvette, France}
			
\author{E.~Frantzeskakis}
\affiliation{Universit\'e Paris-Saclay, CNRS,  Institut des Sciences Mol\'eculaires d'Orsay, 
			91405, Orsay, France}
			
\author{F.~Fortuna}
% \affiliation{Universit\'e Paris-Saclay, CNRS, CSNSM-Centre de Sciences Nucl\'eaire 
%			et de Sciences de la Mati\`ere, 91405, Orsay, France}
\affiliation{Universit\'e Paris-Saclay, CNRS,  Institut des Sciences Mol\'eculaires d'Orsay, 
			91405, Orsay, France}

\author{R.~\surname{Weht}}
\affiliation{Departamento F\'{\i}sica de la Materia Condensada,
      	Comisi\'{o}n Nacional de Energ\'{\i}a At\'{o}mica (CNEA),
    	Avda General Paz y Constituyentes, 1650 San Mart\'{\i}n, Argentina}
\affiliation{Consejo Nacional de Investigaciones Cient\'{\i}ficas
    	y T\'ecnicas (CONICET), Buenos Aires, Argentina}
\affiliation{Instituto de Tecnolog\'{\i}a Sabato, Universidad Nacional de
	  San Mart\'{\i}n - CNEA, 1650 San Mart\'{\i}n, Argentina}
% \email{weht@tandar.cnea.gov.ar}

\author{M.~J.~\surname{Rozenberg}}
\affiliation{Universit\'e Paris-Saclay, CNRS, Laboratoire de Physique des Solides, Orsay 91405, France}

\date{\today }

\begin{abstract}
It was recently discovered that a conductive, metallic state is formed 
on the surface of some insulating oxides. 
Firstly observed on SrTiO$_3$~(001), it was then found in other 
compounds as diverse as anatase TiO$_2$, KTaO$_3$, BaTiO$_3$, ZnO, 
and also on different surfaces of SrTiO$_3$ (or other oxides) with different symmetries.
The spatial extension of the wave function of this electronic state
is of only a few atomic layers.
Experiments indicate its existence is related to the presence of oxygen 
vacancies induced {\it at} or {\it near} the surface of the oxide.
In this article we present a simplified model aimed at describing 
the effect of its small spatial extension
on measurements of its 3D electronic structure 
by angular resolved photoemission spectroscopy (ARPES).
For the sake of clarity, we base our discussion on a simple tight binding scheme plus a confining potential
that is assumed to be induced by the oxygen vacancies. Our model parameters are, nevertheless, obtained
from density functional calculations.
With this methodology we can explain from a very simple concept 
of selective interference the ``wobbling'',
i.e., the photoemission intensity modulation
and/or apparent dispersion of the Fermi surface and spectra 
along the out-of-plane ($k_z$) direction,
and the ``mixed 2D/3D'' characteristics observed in some experiments.
We conclude that the critical model parameters for such an effect are 
the relative strength of the electronic hopping {\it of each band} 
and the height/width aspect ratio of the surface confining potential.
By considering recent photoemission measurements under the light of
our findings, we can get relevant information on the electronic wave functions and of the
nature of the confining potential.
\end{abstract}

\maketitle

%%%%%%%%%%%%%%%%%%%%%%%
\section{Motivation}
%%%%%%%%%%%%%%%%%%%%%%%
Transition metal oxides are fascinating materials which, over the last 
few decades, have been the source of countless surprises,
such as the discovery of high temperature superconductivity in cuprates~\cite{HTc}, 
colossal magnetoresistance in manganites~\cite{manganites},
metal-insulator transitions in vanadates~\cite{ITF}
and, more recently, the resistive switching behavior in various simple and 
complex oxides~\cite{scholarpedia}. 
Understanding many of their properties is at the heart 
of some of the most interesting phenomena and biggest challenges 
in quantum materials science.

Nowadays, modern fabrication methods of thin films
have allowed the realization of heterostructures of these oxides.
These developments bring the promise of observing 
new functionalities which may open the way for novel electronic devices. 
That is the realm of the emergent field of oxide-based electronics~\cite{Ramirez, Cen},
and one of the current focus of major interest is in neuromorphic devices~\cite{jap}.

Among the recent highlights in the study of complex oxide heterostructures
a major breakthrough was the realization, in 2004, of a two dimensional electron gas (2DEG)
at the interface between two well know, insulating and non-magnetic oxides, 
SrTiO$_3$ and LaAlO$_3$~\cite{Ohtomo}.
This triggered a burst of research and new surprises.
For instance, the remarkable finding of a similar 2DEG 
at the bare surface of SrTiO$_3$~(STO)~\cite{nature,natmat}.
This discovery was rapidly followed by similar 
findings in a variety of other insulating materials, 
such as anatase~\cite{2deg-anatase,Moser-anatase},
one of the stable phases of TiO$_2$, 
ferroelectric BaTiO$_3$~\cite{2deg-batio3}, orthorrombic CaTiO$_3$~\cite{2deg-catio3} 
and the strong spin-orbit coupled KTaO$_3$~\cite{2deg-ktao3}. 
In STO, it was also found in other surfaces, particularly on the 
highly anisotropic (110)~\cite{wang_110} and on the highly polar
(111)~ones~\cite{Santander111}. 

Since the initial reports, a great deal of progress was made to investigate 
the origin of the intriguing metallic states appearing at the bare oxide surfaces, 
along with their main properties and on how to improve their quality. 
Experimental evidence strongly suggests that they are created
by the presence of oxygen vacancies ($V_O$) {\it at} or {\it near} 
the surface~\cite{nature, natmat, 2deg-anatase, 2deg-batio3, 2deg-catio3, 2deg-ktao3, 
wang_110, Santander111,roser_svo-sto, Emmanouil, Lomker2017, ZnO}.
Such vacancies seem to have a double key role: 
1) they provide the doping electrons to make the 2DEG 
(i.e. up to two electrons are released by each missing O atom), and, 
2) since they are effectively charged \mbox{+2}, they also produce the confining potential to the 
electron gas~\cite{nature}. 
This basic scenario has been validated by electronic structure calculations~\cite{roser_svo-sto} 
and confirmed experimentally in well controlled synchrotron measurements~\cite{natmat,Emmanouil}, 
where the systematic effect of UV light to create $V_O$ at the surfaces was monitored.
Oxygen vacancies might also play a non-trivial role in the 
electronic and magnetic properties of such 2DEGs, depending on their distance 
to the surface and to the surrounding transition-metal atoms~\cite{Roser-DFT-Ovacs-Magnetism}.

A recent development of great practical relevance that provides further support 
to the aforementioned mechanism, was the discovery that a thin capping layer 
of just a few angstroms thickness of a reducing agent, such as Al or Eu atoms, 
deposited on top of STO produces a very high quality 2DEG~\cite{2deg-batio3,Lomker2017,Emmanouil}. 
In this case, the reducing agent is oxidized by capturing oxygen ions from the STO, leaving 
on its (buried) surface the vacancies that realize the metallic state, and simultaneously creating
a protective capping layer.
This procedure demonstrated its significance by forming an electron gas even
in oxide materials where the oxygen vacancies by themselves are not stable 
or believed to be unable to dope the conduction band~\cite{VanDeWalle2000}
such as the surface of ZnO~\cite{ZnO}, or by realizing a magnetically tunable capping layer,
such as EuO/SrTiO$_3$~\cite{Lomker2017}.
Moreover, this procedure allowed the measurement of a gate-tunable superconducting 
state on the 2DEG state at the STO surface~\cite{shamashis}.

Beyond all these advances, there are several key open questions about these 2DEGs.
For instance, what are the microscopic requirements for an oxide to
realize a 2DEG at its surface when doped with oxygen vacancies. 
Moreover, the actual distribution of those vacancies is currently unknown. 
A related significant conundrum is how to estimate the depth
of the metallic state, i.e. how deep down from the surface it reaches into the material. 
In principle, an ideal two dimensional state would not have any band dispersion 
perpendicular to the surface (henceforth called $z$-axis), and should display tubular 
Fermi surfaces along $k_z$.
%%%
Estimates from density functional theory calculations
indicate that the potential extends only for four or five unit cells beneath the surface~\cite{nature}.

In this sense, assessments of the 2DEG's width from photoemission experiments, 
which can probe the $z$-axis dispersion, remain still controversial. 
Some reports claim that the state is essentially 2D, with the gas being quantum confined 
within about 2~nm and exhibiting subbands~\cite{nature,natmat}. 
However, other photoemission works report a significant  $z$-axis dispersion, 
which was interpreted as an intriguing ``mixed 2D-3D dimensionality'' of the confined state~\cite{plumb}.

The present work will attempt to clarify some of these issues. We shall show that photoemission 
data along the $z$-axis encodes valuable information, which may reveal the qualitative
nature of the confining potential at the surface. The more detailed the photoemission data,
the better estimate of the potential one may obtain. 
In addition, we shall describe how the qualitative features of the $z$-axis
``wobbling'' (the photoemission intensity modulation and/or apparent dispersion) 
can be understood as the effect of the quantum interference of the wave function 
eigenstates (standing waves) along the confinement direction. 
We shall discuss this feature in terms of both the 
spatial extension of the wave function
of the 2DEG and the shape of the confining potential.
Our aim is thus understanding the general effects of quantum confinement on the 
photoemission data along the out-of-plane direction.
%%

%%%
Recently, Moser and coworkers also considered the question of the description 
of the 2DEG along the confinement out-of-plane direction.
They developed a formalism~\cite{moser-2017,moser} based on the damped free electron final state
approximation and the assumption of an exponential confining potential.
Their approach allowed them to extract information on the parameters 
of such exponential confinement potential from comparison to ARPES experimental data. 
They focused their study to InN and anatase TiO$_2$.
%%%

%%%
Here we shall investigate this problem under a different light, namely we incorporate a more 
realistic multi-band structure, including the orbital character and the different effective
masses of the sub-bands, and consider two qualitatively different types of potential wells,
one rectangular and the other wedge-like, to explore which type of confinement 
best represents the more common experimental situations. 
We shall first focus on the paradigmatic case of the SrTiO$_3$(001) surface.
It will be modeled using a tight binding Hamiltonian, with parameters taken from DFT calculations.
We shall show that while the system is actually quantum confined, 
hence strictly non-dispersive along $k_z$,
its electronic structure may appear as having an energy dispersion along $k_z$.
We shall see how this is an \emph{unequivocal} consequence of quantum interference effects
that are different for different sub-bands, and that depend also on the binding energy.
Significantly, we argue that the study of these ``$z$-axis dispersions" can provide 
a clue on the longstanding issue of the lateral 
spatial extension
of the 2DEG state. 
We will then apply our model to anatase.
%%%

%%%
\begin{figure*}[h!t]
  \centering
  \includegraphics[clip,width=0.95\textwidth]{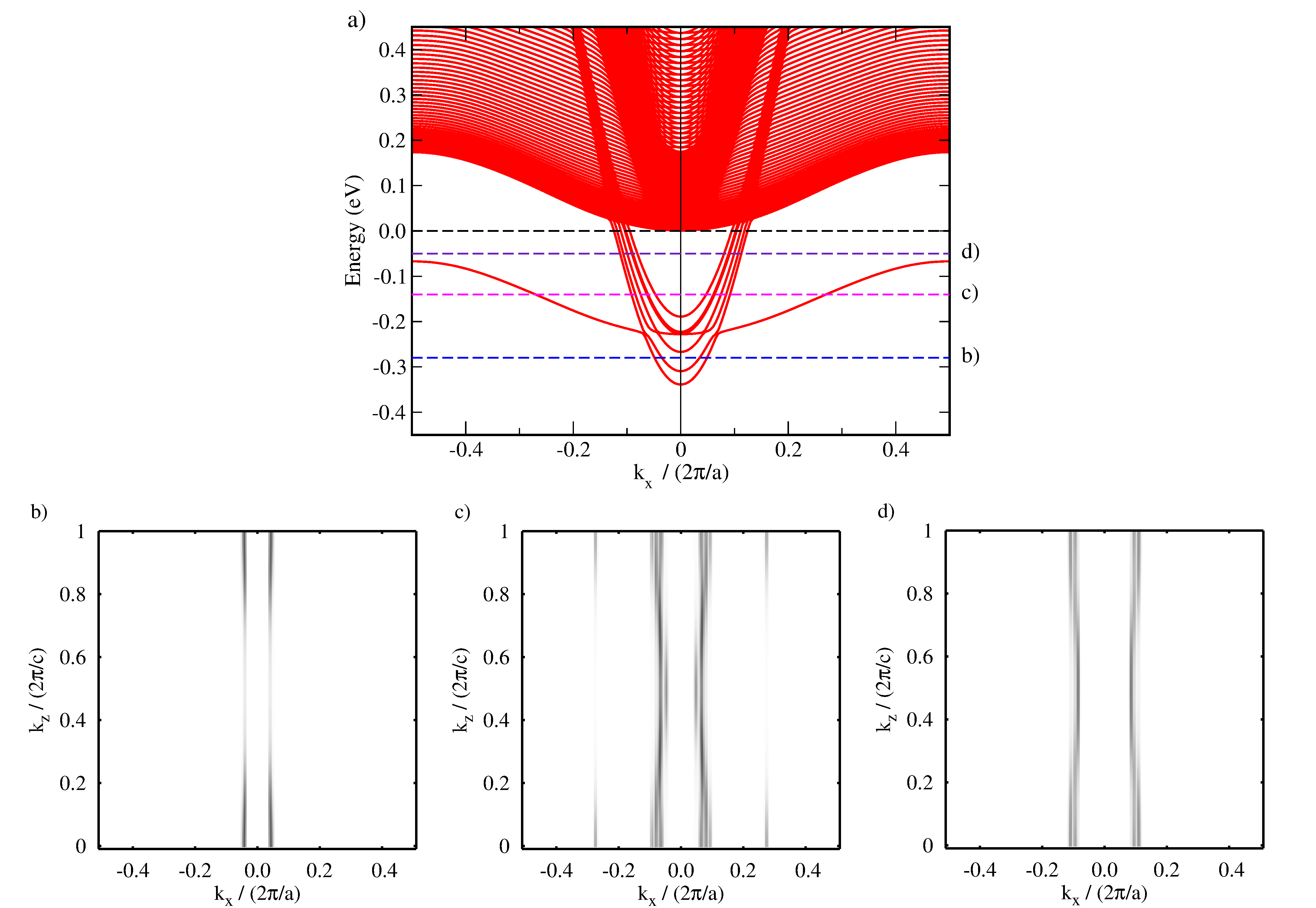}
  \caption{Band structure and Fermi surfaces 
  		  (amplitude squared of the electron wave functions)
  		  for a SrTiO$_3$ slab taking a rectangular confinement potential of 
  		  depth 0.35~eV and a five unit cells width. 
  		  Upper panel, a):~Band structure along a $k_x$ path. 
  		  Different energy levels indicate different amounts of doping
  		  (oxygen vacancies). Explicit values here are:
  		  b)~$-0.28$~eV, c)~$-0.14$~eV, and d)~$-0.05$~eV.
  		  Zero of the energy axis is fixed to the lower conduction state for the bulk.
  		  Lower panel, b) -- d): the corresponding Fermi surfaces in the $k_x$/$k_z$ plane. 
%  		  %%
         }
\label{fig:FS-STO-C}
\end{figure*}
%%%

%%%%%%%%%%%%%%%%%%%%%%%%%%%%%%
\section{Model Hamiltonian for S\lowercase{r}T\lowercase{i}O$_3$}
%%%%%%%%%%%%%%%%%%%%%%%%%%%%%%

Our model for SrTiO$_3$ (STO) is based on a tight binding Hamiltonian,
representing the $t_{2g}$ orbitals, whose hopping parameters we obtained from a fit to a DFT 
band structure calculation --see the Appendix~\ref{Appendix:DFT-STO}.
In STO the valence bands have mostly oxygen character and are completely full,
while the Ti $d$-orbitals occupy the lower conduction bands
and are nominally empty, except when the system is doped. 
In the cubic symmetry, like an undistorted perovskite, Ti $d$-orbitals split, 
due to the octahedral environment, into $t_{2g}$ and $e_g$ levels.
The $t_{2g}$ manifold is usually lower in energy
and closer to the Fermi energy, being 
the first levels to be occupied with doping.
% --Appendix, Fig.~\ref{fig:DFT-STO}.
%%
Mattheiss had already shown how these states form three interpenetrating 
cigar-shape Fermi surfaces, that grow with doping,
along the main crystallographic axes of the cubic Brillouin zone~\cite{Mattheiss}.
The cigar shape emerges because one of the three $t_{2g}$ states 
has a heavy mass (a small curvature in a $E$~vs~$k$ plot) along each direction in  
reciprocal space, while the other two form light bands in the same direction.
In a perfect cubic symmetry, the three $t_{2g}$ states should be
degenerate at $\Gamma$. However, the presence of small tetragonal 
distortions and, even more importantly, the spin-orbit interaction 
slightly breaks this degeneracy. 
Since these effects are relatively small in SrTiO$_3$ and for the sake
of keeping our model as conceptual as possible, 
we shall not consider these details for the moment.
%%%

We complement our STO 2DEG surface model by adding an attractive phenomenological potential,
which is supposed to be experimentally originated by the oxygen vacancies. 
This potential only depends on the perpendicular to the surface $z$-direction, $V(z_i)$, 
where the surface is assumed at $z_0=0$,
and each sub-surface lattice layer is at $z_i$.
We then solve the lattice model and compute all the eigenvalues (energies) 
and eigenvectors (wave functions) of the confined states as a function
of their in-plane momenta $k_{\parallel}$ for each sub-surface layer $z_i$,
and obtain their $k_z$ dependence by Fourier transformation.
Note that, due to the loss of symmetry along $z$, 
$k_z$ is not a good quantum number for the wave functions of the 2DEG. 
Nevertheless, the lack of symmetry does not prevent the Fourier transformation 
along the perpendicular direction, which is merely a change of basis. 
Hence, we use it here to parameterize, as done in experiments, the measured
photon-energy dependence (i.e., the out-of-plane electronic structure) of the
photoemission intensity.
%%%

%%
As mentioned before, we will consider here two prototypical forms for the potential:
a rectangle and a wedge, with their width and depth taken as adjustable parameters.
These shapes are schematic and qualitatively different.
We may assume that the rectangular
shape might be realized in a system where the oxygen vacancies 
are restricted throughout the first few layers below the surface/interface,
as experimentally observed {\textit e.g.} at the interface AlO$_x$/TiO$_2$ 
(see the Supporting Information of Ref.~\onlinecite{2deg-batio3}).
%%%
On the other hand, we may consider the wedge shape as a simplified
representation of an attractive potential that might emerge from a self-consistent 
Poisson-Schr\"odinger calculation~\cite{baumberger}.
As mentioned before, the issue of the actual distribution of oxygen vacancies at, or just beneath,
the surface is still an open problem.

We also consider the related case of anatase TiO$_2$, which has a different crystal symmetry.
The same procedure described before to obtain the tight-binding Hamiltonian 
remains appropriate~\cite{Landmann} 
--see Appendix~\ref{Appendix:DFT-TiO2} for details on the relevant DFT calculations.

\begin{figure*}[b!t]
  \centering
  \includegraphics[clip,width=0.95\textwidth]{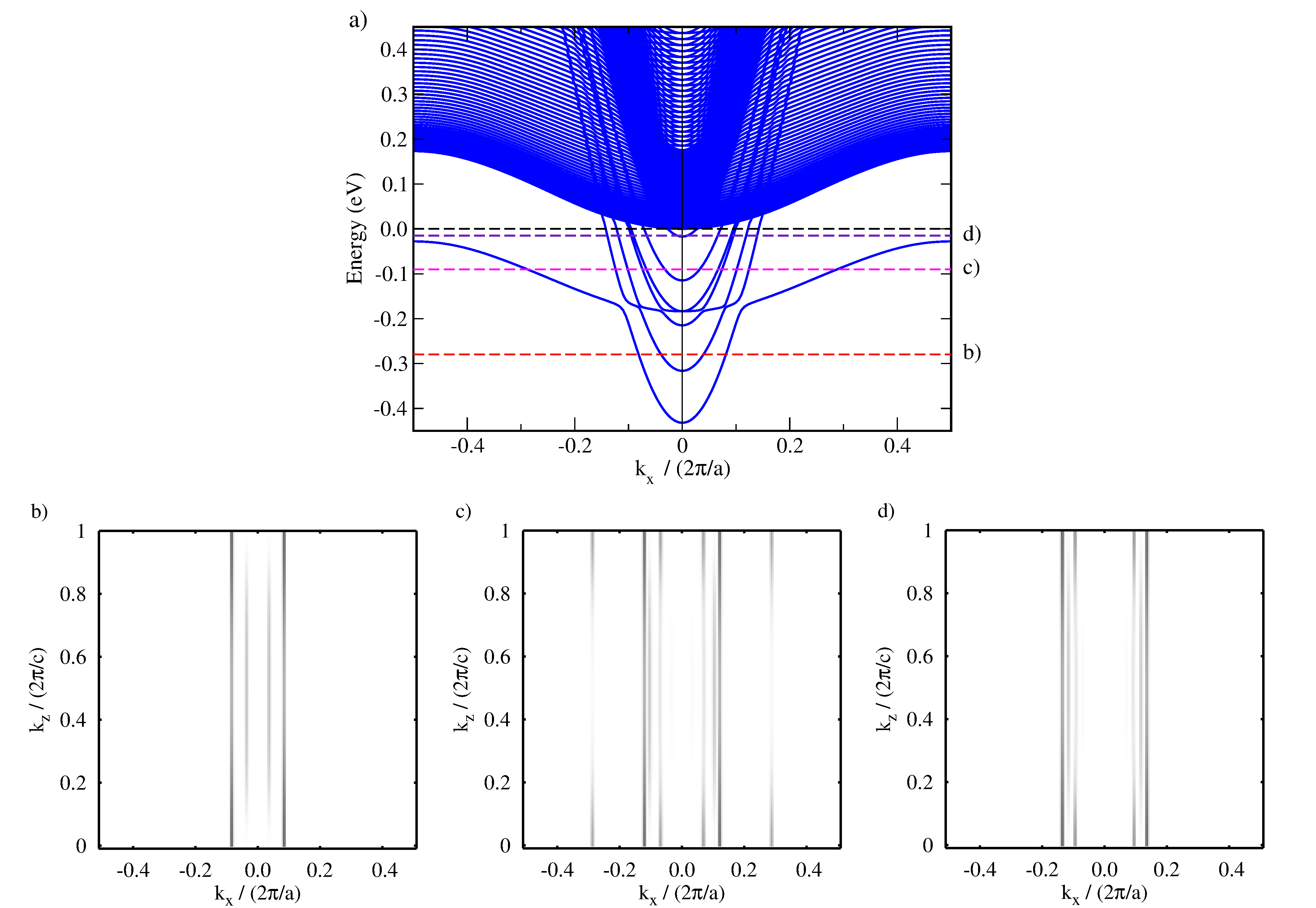}
  \caption{Same as Fig.~\ref{fig:FS-STO-C} but taking a wedge potential of depth 0.50~eV 
  		   and again a five unit cells width. 
  		   Explicit cutting energy levels are:
  		   b)~$-0.28$~eV, c)~$-0.09$~eV, and d)~$-0.016$~eV.
          }
\label{fig:FS-STO-W}
\end{figure*}

%%%%%%%%%%%%%%%%%%%%%%%%%%%%%%
\section{Results and Analysis}
%%%%%%%%%%%%%%%%%%%%%%%%%%%%%%
%%%%%%%%%%%%%%%%%%%%
In the framework of a one-electron description of the electronic structure of the solid,
using Fermi's golden rule and the dipole approximation,
the photo-current $I$
% $I\left(\vec{\epsilon}, E_{\textrm{kin}}, \hbar\omega \right)$ 
of electrons emitted with kinetic energy $E_{\textrm{kin}}$ 
from an initial state $\ket{\psi_i}$ of the 2DEG, 
after excitation with light of energy $\hbar \omega$ and polarization vector $\vec{\epsilon}$,
can be expressed as~\cite{Hufner-book}:
%%%
%%%
\begin{equation}
	I \propto \left|\Braket{\psi_f|\vec{\epsilon}\cdot\vec{\nabla}|\psi_i}\right|^2
			  \times \delta\left(\varepsilon_i + \Phi + E_{\textrm{kin}} - \hbar\omega \right),	
\label{eq:Intensity-PES}
\end{equation}
%%%
%%%
where $\ket{\psi_f}$ is the final state of the optical transition 
and $\Phi$ is the work function. 
Following previous works~\cite{Gadzuk-1974,Mugarza-2003,Pusching-2009,Pusching-2015,moser-2017,moser}, 
we approximate the final state 
as a plane wave in vacuum, with momentum $\vec{k}_f$ 
and damping over a length $\lambda$ from the surface ($z = 0$) towards the bulk, 
to account for the finite escape length of photo-electrons:
%%%
%%%
\begin{equation}
	\psi_f\left(\vec{r}\right) = e^{(i\vec{k}_f - \hat{u}_z/\lambda)\cdot\vec{r}}
	= e^{i\vec{k}_f\cdot\vec{r}} e^{-z/\lambda}, 
\label{eq:FinalState}
\end{equation}
%%%
%%%
where $\hat{u}_z$ is the unit vector along the surface normal.
As the operator $\vec{\nabla}$ is Hermitian, we can let it act on the 
damped free-electron final state of the matrix element, Eq.~\ref{eq:Intensity-PES}, 
giving:
%%%
%%%
\begin{equation}
\begin{split}
	\Braket{\psi_f|\vec{\epsilon}\cdot\vec{\nabla}|\psi_i} & = 
	\vec{\epsilon}\cdot(i\vec{k}_f - \frac{\hat{u}_z}{\lambda})
	\times \\
	& \Braket{e^{(i\vec{k}_f - \hat{u}_z/\lambda)\cdot\vec{r}}|\psi_i}.
\end{split}				 
\label{eq:MatrixElement-FreeElecApprox}
\end{equation}
%%%

%%%
The geometric pre-factor in this expression
depends only on the angle between the light polarization 
$\vec{\epsilon}$ and the direction $\vec{k}_f$ of the emitted electron.
If the energy of the initial state is kept constant, 
as done experimentally when mapping the out-of-plane Fermi surfaces, 
increasing the photon energy to probe the $k_z$ dependence of the electronic structure
will result in a larger final state kinetic energy and, consequently, 
a larger final state momentum $\vec{k}_f$.
Hence, far from any resonances, the main effect 
of increasing the photon energy will simply be 
to change smoothly the amplitude of the geometric pre-factor and 
to increase the size of reciprocal space probed by photoemission. 
Thus, we can approximate the photo-current 
as~\cite{Pusching-2009,Pusching-2015,moser-2017,moser}:
%%%
\begin{equation}
	I(\vec{k}_f) \propto \left| \int_0^{\infty}  
	e^{i \vec{k}_f \cdot \vec{r}} 
	e^{-z / \lambda}
	\psi_i (\vec{r}) 
	d\vec{r}
	\right|^2.
\label{eq:Intensity-FreeElecApprox}
\end{equation}
%%%

%%
In other words, within the free-electron final-state approximation,
the photocurrent is proportional to the amplitude squared of the
Fourier transform of the damped initial state wave-function. 
The inclusion of a finite damping length complicates a bit the analysis 
without altering the qualitative results~\cite{moser}
--see the Appendix, Fig.~\ref{fig:Damping}. 
%%%
As we wish to focus on the essential consequences of quantum interferences 
induced by confinement on the electronic structure along $k_z$,
we will henceforth neglect the effects of finite photo-electron escape length and
dipole matrix elements. 
In this way, the importance of the more superficial states will be immediately evident. 
Thus, in what follows, we will simply model the photoemission intensity as:
%%%
\begin{equation}
	I(\vec{k}) \propto \left| \tilde{\psi}_{i}(\vec{k}) \right|^2,
	\label{eq:Intensity-FourierTransform}
\end{equation} 
\textit{i.e.,} as the modulus squared of the initial state's Fourier transform 
$\tilde{\psi}_{i}(\vec{k})$~\cite{Mugarza-2003,Pusching-2009,Pusching-2015}.
%%%

%%%
In Fig.~\ref{fig:FS-STO-C} and \ref{fig:FS-STO-W} 
we show our modeling results for SrTiO$_3$(001) using both 
a rectangular and a wedge confining potential, respectively. 
In the corresponding upper panels we show the band structures of 
our system along the $k_x / k_y$ coordinates --the $x$ and $y$ directions are 
degenerate by symmetry, we will note them $k_{\parallel}$
when the states are degenerated (or almost degenerated) in the whole plane.
The multiple bands above the zero energy level (taken here as the lowest
conduction state in the bulk) are bulk states.
Their multiplicity comes from the real space representation of our slab along $z$,
that will become a shaded region in a true semi-infinite system.
Bands below zero energy are the sub-bands, pulled down by the confining potential.
The number of sub-bands and their energy positions will depend  
on the shape and depth of the confining potential,
while their curvatures (i.e. effective masses) are essentially given 
by the respective bulk bands from which they come from. 
In both, rectangular and wedge-like potential cases we chose the parameters of the wells 
to pull down a similar number of bands, five light and one heavy. 
The relative asymmetry in the number of bands pulled down can be most easily understood
in terms of the perpendicular effective mass,
%%%
defined as proportional to the inverse of the hopping integral along the out-of-plane direction.
%%%
By the $t_{2g}$ symmetry the light
bands in $k_x / k_y$ will have a heavy $z$-mass and vice-versa. Hence light bands
in the in $k_x / k_y$ plane are relatively more easily pulled down~\cite{nature}. 
%%%

In the lower panels of Figs.~\ref{fig:FS-STO-C} and \ref{fig:FS-STO-W} 
we plotted the corresponding Fermi surfaces
(visualized as the amplitude squared of the electron wave functions), 
on the $k_x / k_z$ plane, for some given energy levels indicated by the dashed horizontal lines 
in the upper panels. This energy levels would correspond to some Fermi energies
given by the occupation of the sub-bands.
In an actual experiment this level will depend on total doping induced by the 
dosing of oxygen vacancies, or eventually by an applied gate potential. 
%%%

%%%
\begin{figure}[bt]
  \centering
  \includegraphics[clip,width=\columnwidth]{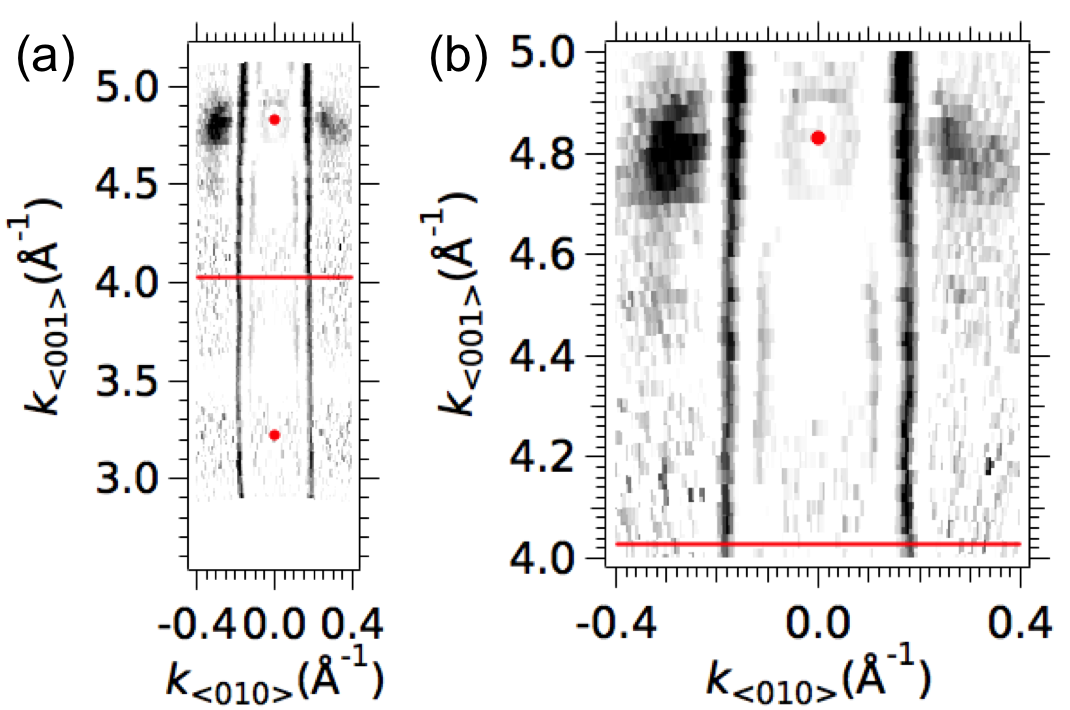}
  \caption{
  		  (a)~ARPES Fermi surface of EuO(1~ML)/SrTiO$_3$ along the $k_z$ axis, 
  		   acquired by varying the photon energy between 30~eV and 97~eV in steps of 1~eV 
  		   using linear vertical photon polarization. 
  		   An inner potential of 16~eV was used in the calculation
  		   of the out-of-plane momentum~\cite{Lomker2017}.
           The red discs and lines mark, respectively, the bulk $\Gamma$ points
           and the border of the bulk Brillouin zone. 
           (b)~Zoom over the upper part of the Fermi surface in panel~(a).
           All data in this figure were acquired at 15~K, 
           and correspond (for better signal-to-background ratio) to the
           negative part of the second derivative of the raw data.
          }
\label{fig:ESTO}
\end{figure}
%%%

%%%
\begin{figure*}[b!ht]
  \centering
  \includegraphics[clip,width=1.00\textwidth]{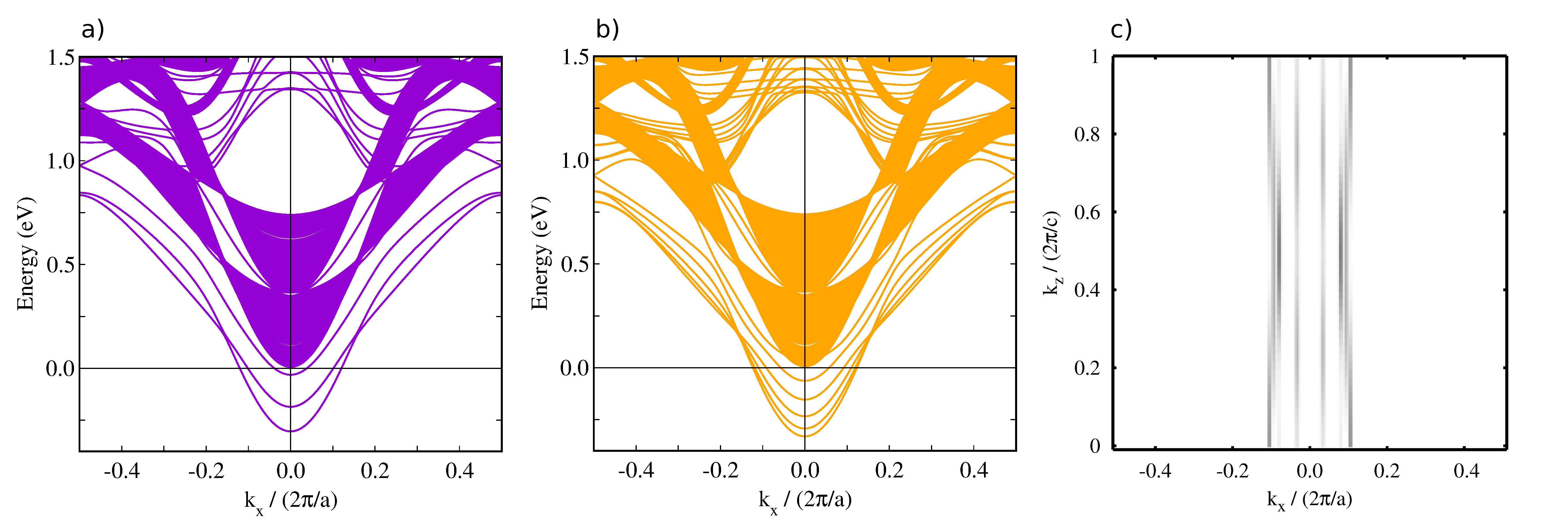}
  \caption{Band structures and Fermi surface 
  		   (amplitude squared of the electron wave functions)
  		   for an anatase TiO$_2$ slab with a rectangular confining potential of
	   	   depth 0.35~eV and two different widths: 
	   	   a)~5~TiO$_2$ layers, 
	   	   b)~10~TiO$_2$ layers.
	   	   c)Corresponding Fermi surface for the case with 10 layers and $E_F =-0.10$~eV.
          }
\label{fig:FS-Anatase}
\end{figure*}
%%%

%%%
We first focus on Fig.~\ref{fig:FS-STO-C}. At first glance,
some of the Fermi surface sheets appear as weakly dispersive in $k_z$, 
corresponding to quasi-2D tubular Fermi surfaces. 
However, upon closer inspection we can notice quite interesting effects.  
In fact, while each of the Fermi sheets is strictly non dispersive,
the amplitude squared of the electron wave functions
show an intensity modulation in $k_z$, due to interference effects of each wave function
in the confining potential well, that gives the impression of an incipient Fermi-surface dispersion, 
most visible in panels c) and d). 
In contrast, such an effect is less apparent in panel b). 
The reason can be traced back to the number of bands
that are crossing the Fermi energy, their separation and orbital characters. 
We see that there are only two bands crossing $E_F$ in panel b), 
while a bundle of at least five bands crosses it in panels c) and d). 
This is an interesting observation, since the resolution in ARPES experiments
of all possible sub-bands predicted by calculations 
remains an experimental challenge~\cite{baumberger}.
Thus, as shown by our simple modeling, even a seemingly small Fermi-surface dispersion
along $k_z$ in the experimental data would be actually due to the combined effects,
on \textit{multiple sub-bands}, of intensity modulations of their electron wave-functions 
and experimental resolution broadening. 
Hence, a detailed analysis of the out-of-plane experimental data along the lines exposed here
would provide a means to probe, or even unveil, 
different sub-bands {\it beyond} the nominal experimental resolution. 
%%%

%%%
Another interesting feature is that the out-of-plane intensity modulation of a given
state depends on its ``effective mass'' 
(inverse of the hopping integral) 
along $k_z$.
For instance, the modulation of the outmost Fermi surface in Fig.~\ref{fig:FS-STO-C}~(c)
is stronger than the one in Fig.~\ref{fig:FS-STO-C}~(b). 
In fact, as seen in Fig.~\ref{fig:FS-STO-C}~(a), while the former derives from a
$d_{yz}$ state of heavy mass in $k_x / k_y$ but light mass in $k_z$,
the Fermi surface in Fig.~\ref{fig:FS-STO-C}~(b) stems from the first couple of $d_{xy}$ subbands,
which are light in $k_{\parallel}$ but heavy in $k_z$. 
Thus, a light out-of-plane mass implies a stronger quantum modulation 
of its Fermi-surface intensity. 

It is noteworthy that the results of Fig.~\ref{fig:FS-STO-C}-c) 
resemble qualitatively the data reported by Plumb et. al.~\cite{plumb} for the out-of-plane
Fermi surface of the 2DEG at the SrTiO$_3$(001) surface.
Thus our analysis puts under a clearer light the puzzling interpretation of ``mixed 2D-3D dimensionality''
advanced by those authors, 
replacing it into the more precise context of a collection of tubular Fermi surfaces
with different intensity modulations along $k_z$, all of which create the impression
of a 3D Fermi surface dispersing in the out-of-plane direction for the in-plane heavy sub-band.
%%%

%%%
To gain further insight we now turn to the wedge potential case
that is shown in Fig.~\ref{fig:FS-STO-W}.
The main qualitative observation upon comparison to the previous case is
that the intensity modulation is also very strong, however the ``wobbling''
dispersive effect along $k_z$ is now significantly less pronounced. 
%%%
The reason for the less pronounced wobbling can be traced to the fact that, 
unlike the case of the rectangular potential, 
now the light sub-bands appear more separated from each other, 
as can be seen in panel Fig.~\ref{fig:FS-STO-W}~(a).
This is due to the nature of the wedge potential, which has the form $V(z_i)$, 
and therefore acts as a different crystal field for each sub-surface lattice layer $z_i$.
This ``layer-dependent crystal field'' fully breaks the quasi-degeneracy of the rectangular
potential case, where all sublayers are subject to the same potential $V_o$ up to the well thickness. 
%%%

%%%
This is a very interesting and rather surprising observation, since one may naively 
expect that a self-consistent Poisson-Schr\"odinger treatment of the surface 
would be more appropriate~\cite{baumberger}. 
However, in most data available for the $k_z$ dispersion
of 2DEGs in oxides, one does observe a rather smooth but clear wobbling of the dispersion. 
We can illustrate this with previously unpublished ARPES data in SrTiO$_3$,
shown in Fig.~\ref{fig:ESTO}. The excellent quality of the data is due to
the surface preparation method, which consisted in capping the bare STO surface 
with an atomically thin layer of pure Eu~\cite{2deg-batio3,Lomker2017}.
Here, we observe a clear modulation effect,
especially for the outermost Fermi surface (corresponding to the in-plane heavy subband),
in good qualitative agreement with the results of Fig.~\ref{fig:FS-STO-C}(c).
%%%
So, a qualitative comparison with the simulations insinuates 
that the experimentally observed wobbling might be related 
to a more rectangular-like attractive potential acting on the first few sub-surface layers,
suggesting that the oxygen vacancies are distributed within those layers.  
However, we also note that other details seem to be better captured by the wedge potential. 
For instance, in Fig.~\ref{fig:FS-STO-W}(b) we see that the quantum interfere modulation 
picks up a stronger intensity near the $\Gamma$ point ($k_z = 0$), 
similarly as in the experimental data.
%%%

%%%
In this sense, note that previous experimental studies of TiO$_2$-anatase 
showed that the oxygen vacancies responsible for the formation of the 2DEG and the 
confining potential well are approximately homogeneously distributed 
over around 1nm below the surface (see Ref.~\onlinecite{2deg-batio3}, in particular
Supplementary Figure~5 and related discussion). 
Thus, these vacancies would form a positively charged slab of finite thickness, 
comparable to the spatial extension of the 2DEG's electronic wave-function, 
rather than a strictly 2D charged plane at the surface. 
While the latter would produce a wedge-like confining potential, 
the former would actually produce a quadratic (parabolic) confining potential 
in the region where the vacancies are distributed, 
followed by a linear wedge-like potential beneath such a region towards the bulk,
and bound by an infinite potential barrier at the surface. 
Thus, in the region where the 2DEG states are located, 
the potential well due to vacancies would be more ``rectangular'',
while it might become ``wedged'' beneath.
%%%

%%%
We leave the precise quantitative evaluation of the confining well that can best describe
the experimental data for future research. 
Here, our goal is to emphasize the qualitative implications of quantum-mechanical confinement, 
in the simplest possible context, for the understanding 
of the out-of-plane ARPES data of 2DEGs in oxides.
%%%%%

%%%
To conclude our study, we shall consider the case of a different oxide,
TiO$_2$ anatase~(001),  which in contrast to STO has a tetragonal crystal symmetry.
We also take this case to explore the effect of the depth of the well.
 
The conduction bands of bulk anatase TiO$_2$ also have $t_{2g}$-type character, like in STO.
However, in contrast to the previous compound, the tetragonal symmetry breaks down 
their degeneracy and the orbitals separate in two groups, 
with one and two bands per atom in the unit cell. 
Moreover, also here the lower conduction band in the bulk has the complementarity 
of being light in the $k_x / k_y$ plane, but heavier along $k_z$
--~see the Appendix, Fig.~\ref{fig:DFT-TiO2}.
%%%

In Fig.~\ref{fig:FS-Anatase} we show the calculated band structures for two rectangular
confining potentials, both of depth 0.35~eV but with different widths along the $z$-axis.
Fig.~\ref{fig:FS-Anatase}(a) shows the case of five TiO$_2$ layers,
while Fig.~\ref{fig:FS-Anatase}(b) considers the case of ten TiO$_2$ layers.
Fig.~\ref{fig:FS-Anatase}(c) shows the results for the $k_z$ axis dispersion 
for the thicker well. We observe that all the previously mentioned features arise, 
namely a rather smooth wobbling of the dispersion 
for an energy cut that goes through a tightly packed bundle of sub-bands.
Such a bundle would unlikely be experimentally resolved along the $k_{\parallel}$ direction,
but we see here that the $k_z$ data can provide hints to its existence.

%%%
\section{Conclusions}
%%%
We have considered a simple tight binding Hamiltonian, with realistic
hopping parameters derived from DFT, to build a model of a
quasi semi-infinite lattice with a confining potential at the surface. 
We analyzed the qualitative differences for different shapes of the surface
potential well, relevant for ARPES experiments of 2DEGs at the surface of transition
metal oxides.
Our analysis allowed us to provide a qualitative understanding of the
interference phenomena along the $z$-axis direction originated
in the few-layers quantum confinement of the 2DEGs.
Our results provided clarification of a puzzling ``mixed dimensionality''
state reported in SrTiO$_3$ experiments. By the comparison of different
confining potential shapes we argue that, contrary to naive expectations,
the rectangular well seems to better represent the more common 
experimental situation. 
This underlines the need to achieve a better understanding
of the (self-)organisation of the oxygen vacancies that originate the 2DEGs.
We also discussed how the observation of a smooth wobbling of the bands 
along the $z$-axis, which so far has been hardly studied, may provide valuable indication of
the presence of sub-bands that are tightly packed, beyond the experimental 
resolution limit. In this spirit, we may envision that a careful fit of the ARPES data 
along the $z-$axis may provide detailed information on the shape of the confining potential and, 
henceforth, on the reorganization of the oxygen vacancies beneath the surface.
This appears as an exciting future challenge within our reach. 

\newpage
\acknowledgments
{
This work was supported by public grants from the French National 
Research Agency (ANR), project LACUNES No ANR-13-BS04-0006-01, 
the ``Laboratoire d'Excellence Physique Atomes Lumi\`ere Mati\`ere''
(LabEx PALM projects ELECTROX and 2DEG2USE) overseen by
the ANR as part of the ``Investissements d'Avenir'' program 
(reference: ANR-10-LABX-0039), and the CNRS-CONICET 2015-2016 
collaborative project AMODOX (project number: 254274).
R.W. gratefully acknowledges 
CONICET and ANPCyT (Grant No. PICT-2016-0402), Argentina,
for partial support 
and Dario Carballido (FRED-CNEA) for computing assistance.
}

%%%%% APPENDIX
\appendix
%%%%%%%%%%%%%%%

%%%%%%%%%%%%%%%%%%%%%
\section{DFT and tight-binding model for SrTiO$_3$}
\label{Appendix:DFT-STO}
%%%%%%%%%%%%%%%%%%%%%

\begin{figure*}[h!tb]
  \begin{center}
	  \includegraphics[width=0.90\textwidth]{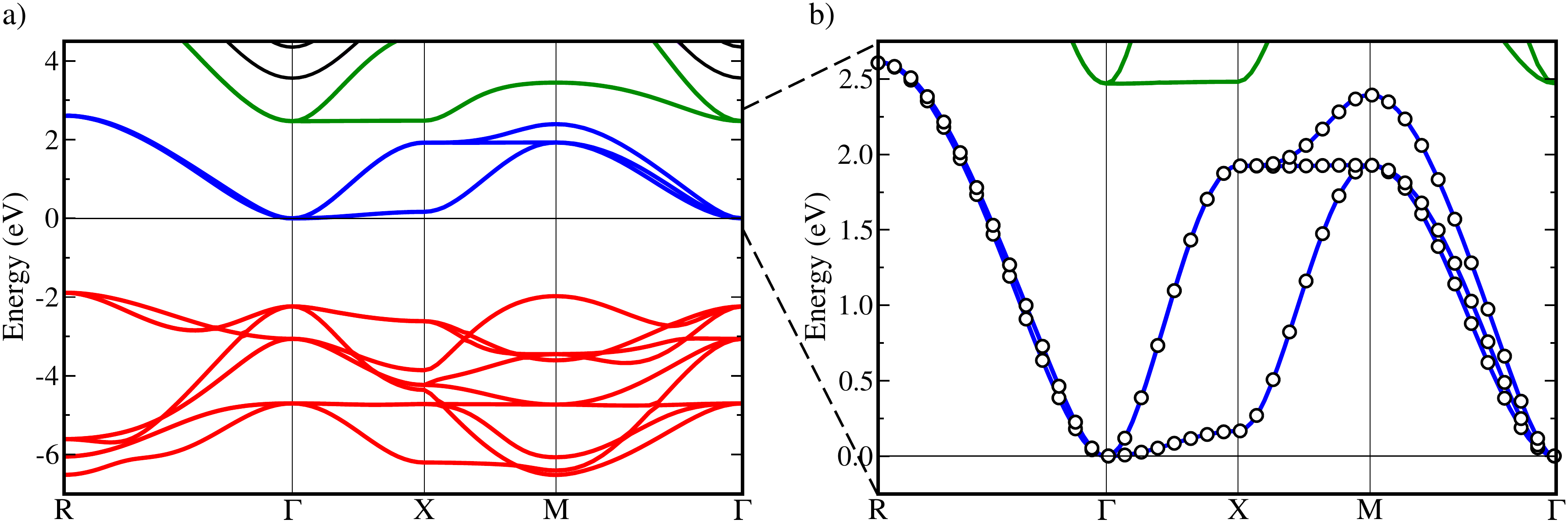}
  \end{center}
	\caption{
	a) Standard DFT bands for SrTiO$_3$.
	The calculation method is DFT, using the code Wien2k. 
	Red lines correspond to bands with mostly oxygen character.
	Blue and green bands to the $t_{2g}$ and $e_g$ manifolds, respectively.
	b) Wannier fitting to the $t_{2g}$ bands (black circles). 
          }
    \label{fig:DFT-STO}
\end{figure*}

%%
DFT calculations of SrTiO$_3$ band structure were done 
starting from a standard all electron density functional
calculation, using in our case the Wien2K code~\cite{wien2k}.
The obtained bands are shown in Fig.~\ref{fig:DFT-STO}(a).
We then projected the $t_{2g}$ states to get the tight binding model parameters
for SrTiO$_3$.
We adopted Maximally Localized Wannier functions, following the prescription
by Marzari et al.~\cite{MLWF} and using the {\small \rm WIEN2WANNIER} program~\cite{Kunes}, 
as shown in Fig.~\ref{fig:DFT-STO}(b).
With those hopping parameters, we construct a tight binding lattice that
on the $x/y$ plane keeps the infinite periodicity of the crystal, while along 
the $z$-direction we build a thick, real space slab of up to 200 layers.

The lattice sites correspond actually to the Ti positions while 
the effects of the oxygen orbitals are implicitly taken into account 
in the computation of the hopping integrals between the Wannier functions.
We did not do any mass renormalization on the hopping parameters.

%%%%%%%%%%%%%%%%%%%%%
\section{DFT model for TiO$_2$ anatase}
\label{Appendix:DFT-TiO2}
%%%%%%%%%%%%%%%%%%%%%

\begin{figure*}[h!tb]
  \begin{center}
	  \includegraphics[trim=0cm 1cm 0cm 0cm, width=0.7\textwidth]{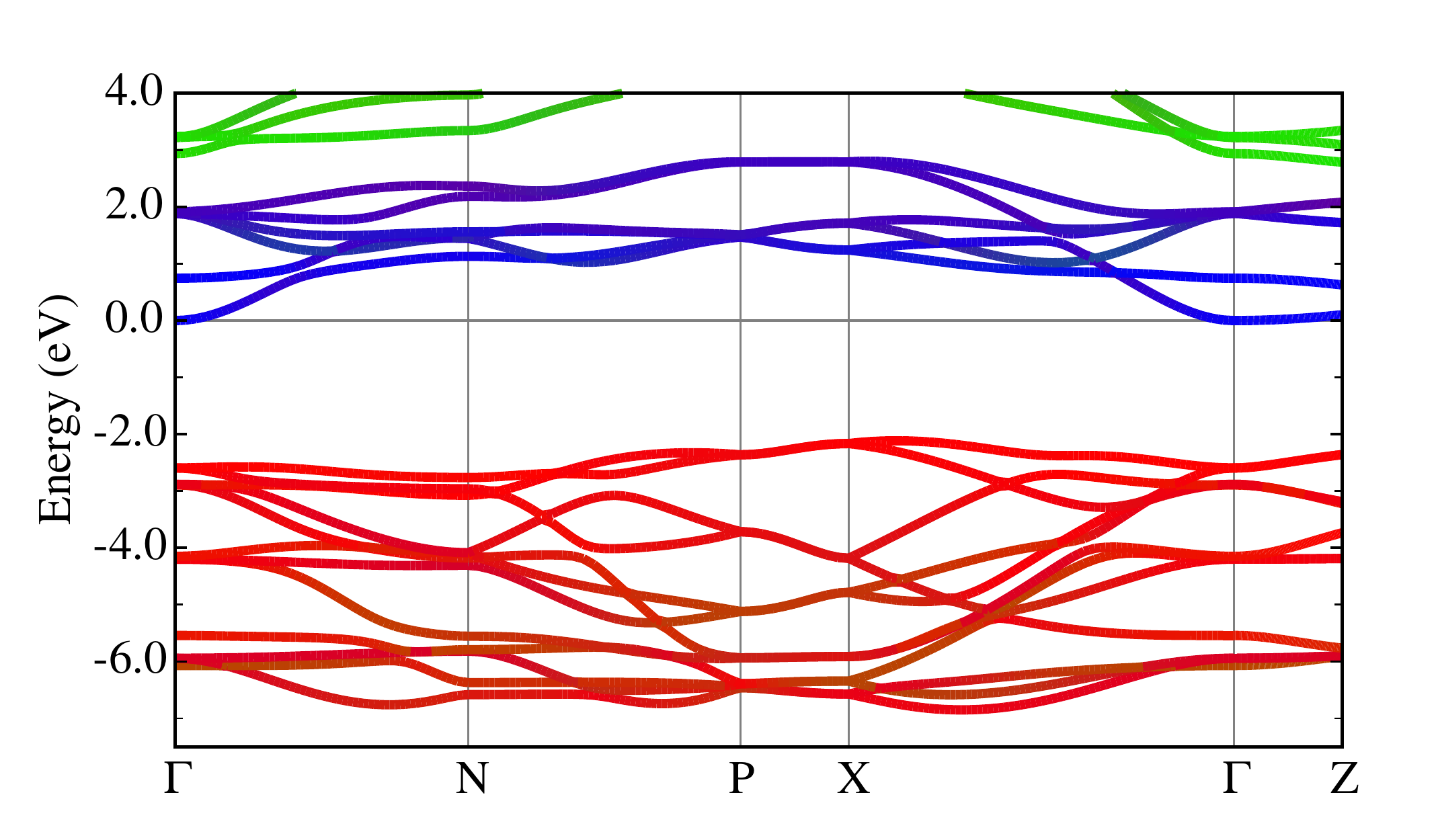}
  \end{center}
	\caption{DFT band structures for Anatase. 
	Blue character means t$_{2g}$ orbitals, 
	green e$_g$ and red oxygen orbitals.
          }
    \label{fig:DFT-TiO2}
\end{figure*}

DFT calculations for TiO$_2$ anatase were performed following the same procedure as for SrTiO$_3$.
The obtained bands are shown in Fig.~\ref{fig:DFT-TiO2}.
Note that anatase has two Ti atoms in the unit cell. Therefore, our Wannier
basis set has six orbitals. We do consider in this case a mass renormalization factor
of $0.6$, but its effect on the reported Fermi surfaces is rather minor.

%%%%%%%%%%%%%%%%%%%%%
\section{Effect of the finite photoelectron escape length}
\label{Appendix:Damping}
%%%%%%%%%%%%%%%%%%%%%

\begin{figure*}[h!tb]
  \begin{center}
	  \includegraphics[width=.95\textwidth]{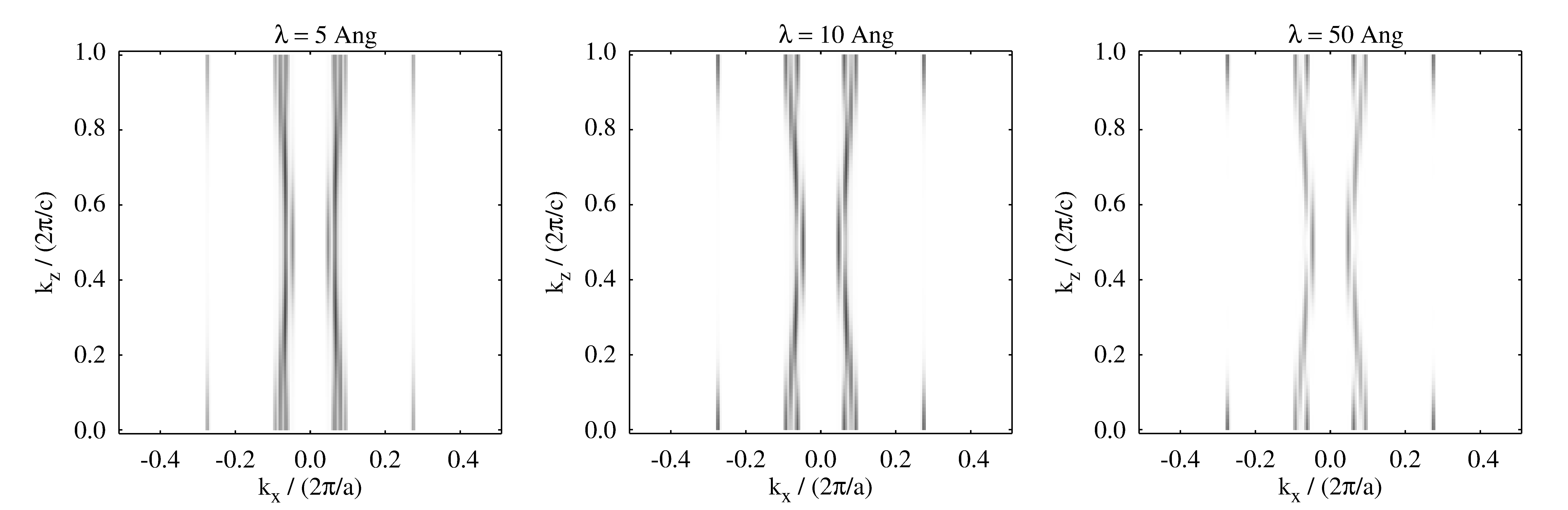}
  \end{center}
	\caption{Effect of the damping parameter $\lambda$ (Eq.~\ref{eq:Intensity-PES}) 
			 on the plot of Fig.~\ref{fig:FS-STO-C}(c) for (from left to right)
			 $\lambda = 5$~\AA, 10~\AA, and 50~\AA.
          }
    \label{fig:Damping}
\end{figure*}

To quantify the effect of the finite inelastic mean free path of photoelectrons
(equation~\ref{eq:Intensity-PES}) compared to the case where such a damping is neglected, 
we show in Fig.~\ref{fig:Damping} the effect of different photoelectron escape lengths 
$\lambda = 5, 10,\textrm{ and } 50$~\AA~on the simulation of the out-of-plane Fermi surface
for the case of a 2DEG confined by a rectangular potential at the surface of SrTiO$_3$
for a doping corresponding to Fig.~\ref{fig:FS-STO-C}(c).
As seen by a comparison of those two figures, while the damping accentuates 
the $k_z$ modulation of the intensity, the effect caused by the interference
of the electron wave-function in the potential well remains the essential feature
giving raise to the apparent dispersion of the out-of-plane Fermi surface. 

%%%%%%%%%%%%%%
%%%%%%%%%%%%%%%

%%%%%%%%%%%%%%%
%%%%%%%%%%%%%%%

%%%%%%%%%%%%%%%
%%%%%%%%%%%%%%

%%\balance

\end{document}